\documentclass[aps, pra, preprint, amsmath, amssymb, superscriptaddress]{revtex4-2}

\usepackage{graphicx}
\usepackage{dcolumn}
\usepackage{bm}
\usepackage{xcolor}
\usepackage[onehalfspacing]{setspace}
\usepackage[colorlinks=true,
            linkcolor=blue,
            citecolor=blue,
            urlcolor=blue]{hyperref}

\begin{document}

\title{A dipolar Bose-Bose mixture of Dysprosium isotopes with controllable interspecies interactions}

\author{M. D\"{u}rbeck}
\thanks{These authors contributed equally to this work.}
\affiliation{Fritz-Haber-Institut der Max-Planck-Gesellschaft, Faradayweg 4-6, 14195 Berlin, Germany}

\author{L. Reihs}
\thanks{These authors contributed equally to this work.}
\affiliation{Fritz-Haber-Institut der Max-Planck-Gesellschaft, Faradayweg 4-6, 14195 Berlin, Germany}

\author{J.P. Marulanda-Serna}
\affiliation{Fritz-Haber-Institut der Max-Planck-Gesellschaft, Faradayweg 4-6, 14195 Berlin, Germany}

\author{B. Choudhari}
\affiliation{Fritz-Haber-Institut der Max-Planck-Gesellschaft, Faradayweg 4-6, 14195 Berlin, Germany}

\author{J. Seifert}
\affiliation{Fritz-Haber-Institut der Max-Planck-Gesellschaft, Faradayweg 4-6, 14195 Berlin, Germany}

\author{N. Werum}
\affiliation{Fritz-Haber-Institut der Max-Planck-Gesellschaft, Faradayweg 4-6, 14195 Berlin, Germany}

\author{G. Meijer}
\affiliation{Fritz-Haber-Institut der Max-Planck-Gesellschaft, Faradayweg 4-6, 14195 Berlin, Germany}

\author{G. Valtolina}
\email{valtolina@fhi-berlin.mpg.de}
\affiliation{Fritz-Haber-Institut der Max-Planck-Gesellschaft, Faradayweg 4-6, 14195 Berlin, Germany}
\affiliation{Humboldt Universit\"{a}t zu Berlin, Newtonstrasse 15, 12489 Berlin, Germany}

\date{\today}

\maketitle

\section*{Abstract}
We report on the realization of a quantum-degenerate Bose-Bose mixture of $^{162}$Dy and $^{164}$Dy. Owing to the near-identical mass and polarizability of the two isotopes, the mixture thermalizes efficiently, with evaporation trajectories closely following those of the single-isotope case. Using a broad interspecies Feshbach resonance, we explore a miscible-immiscible transition between the two Bose-Einstein condensates. The tunability of the interspecies interaction, combined with the large magnetic dipole moment of Dy, makes this platform well suited for exploring dipolar effects in ultracold mixtures, including multi-component supersolidity.

\section{Introduction}
Ultracold gases of highly magnetic lanthanide atoms, such as dysprosium (Dy) and erbium (Er), have recently emerged as an ideal system for the investigation of quantum gases with long-range and anisotropic dipole-dipole interactions. The large magnetic dipole moment in Dy and Er stems from the large orbital angular momentum of their electronic ground state. In Dy, the ground state has a total angular momentum $J=8$, which results in a magnetic dipole moment of $10\ \mu_B$, where $\mu_B$ is the Bohr magneton. The control over large magnetic dipole-dipole interactions in lanthanide quantum gases has recently enabled the observation of dipolar supersolids, a phase of matter characterized by the simultaneous presence of phase-coherence and density modulation \cite{chomaz_dipolar_2022,tanzi_observation_2019,bottcher_transient_2019,chomaz_long-lived_2019}. More exotic regimes of supersolidity have been predicted in multi-component dipolar condensates \cite{trautmann_dipolar_2018,bland_alternating-domain_2022, Bisset_Quantum_droplets_2021, scheiermann2023, Arazo2023, AuChen2022}.
However, spinor mixtures of lanthanides suffer from strong inelastic losses due to dipolar relaxation \cite{Burdick_Fermionic_suppression_2015,claudeopticalmanipulationspinstates2024,Lecomte_production_2025}.\newline
In this work, we realize a stable quantum degenerate mixture of dipolar atoms consisting of the isotopes $^{162}$Dy and $^{164}$Dy, the two most abundant bosonic isotopes of Dy. Since both isotopes are trapped in their lowest spin state, dipolar relaxation is naturally suppressed. This allows us to realize a mixture with good collisional properties that can be further controlled via convenient interspecies Feshbach resonances. Moreover, their combination constitutes the most dipolar mixture realized to date, establishing Dy isotope mixtures as an ideal platform for investigating multi-component dipolar gases and new regimes of dipolar superfluids and supersolids.

\begin{figure}
 \centering
        \includegraphics[width=1\textwidth]{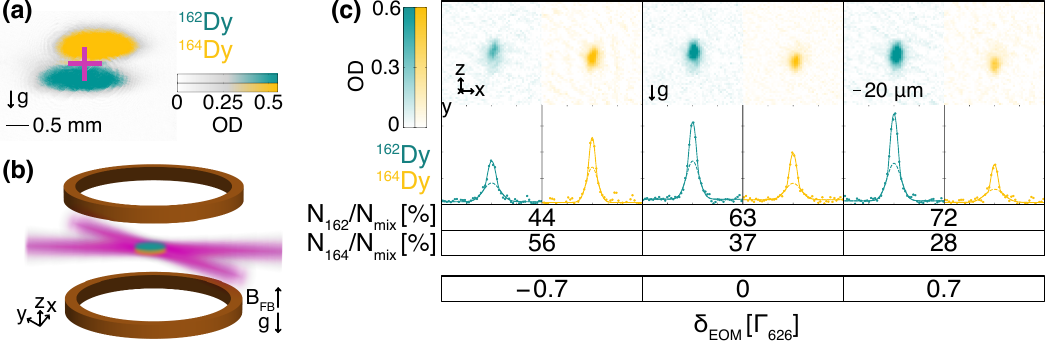}
 \caption{(a) Absorption images  of $^{162}$Dy and $^{164}$Dy in the DI-MOT at $\delta_\mathrm{EOM}=-3.0\ \Gamma_{626}$,  with $\Gamma_{626}$ being the natural linewidth of the 626-nm  transition. The vertical displacement allows us to control the population imbalance in the crossed ODT, whose position is marked by a purple cross. (b) Sketch of the experimental apparatus showing the atoms being loaded from the DI-MOT into the crossed ODT. (c) Left:  At negative detuning  $\delta_\text{EOM}=-0.7\ \Gamma_{626}$, the loading of $^{164}$Dy into the ODT is favored over $^{162}$Dy. Here, we realize a Bose-Einstein condensate (BEC) of $^{162}$Dy and $^{164}$Dy consisting of  $N_{162} = 1.4\cdot10^{4}$ and $N_{164} = 1.8\cdot10^{4}$, with a condensed fraction of around 25$\%$ and 40$\%$ respectively. $N_{162}$ and $N_{164}$ are the atom numbers of $^{162}$Dy and $^{164}$Dy respectively. Center: for $\delta_\text{EOM}=0$ the mixture is composed of a BEC of $N_{162} = 2.2\cdot10^{4}$ and $N_{164} = 1.3\cdot10^{4}$, with a condensed fraction of around 50$\%$ and 20$\%$ respectively. Right: for $\delta_\mathrm{EOM}=0.7\,\Gamma_{626}$, we favorably load $^{162}$Dy into the ODT and realize a mixture of $N_{162} = 2.6\cdot10^{4}$ and $N_{164} = 1.0\cdot10^{4}$, with a condensed fraction of around 55$\%$ and 10$\%$ respectively. Additionally, we report the population fraction of each isotope, where $N_\mathrm{mix}=N_{162}+N_{164}$ refers to the total atom number of the mixture.}
\label{fig1}
\end{figure}

\section{Experimental sequence}
\begin{figure}
 \centering
        \includegraphics[width=1\textwidth]{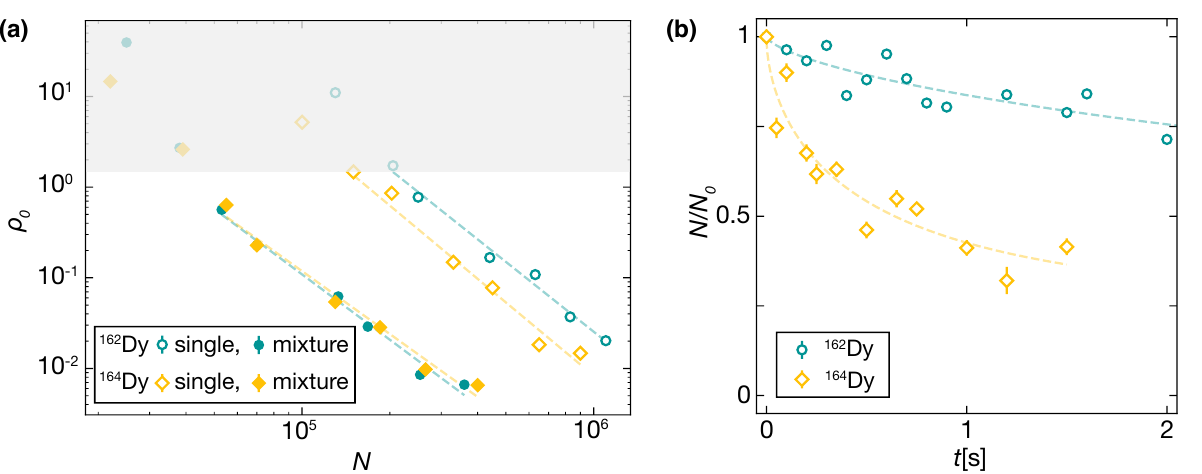}
 \caption{Evaporative cooling trajectories for single-species and dual-species operation at a magnetic field of 9.7\,G. (a) Peak PSD $\rho_0$ as a function of atom number $N$ during forced evaporation of $^{162}$Dy (green) and $^{164}$Dy (orange) in single-species (hollow) and mixture (solid) experiments. We extract $\gamma$ as the slope of the respective linear fits (dashed lines). The gray band marks the threshold PSD for the onset of condensation (PSD $\sim 1$). The near-identical slopes in the single-isotope and mixture configurations indicate that the interspecies interactions do not greatly affect the evaporation efficiency at a magnetic field of 9.7\,G.  b) Lifetime measurements of $^{162}$Dy BEC ($\tau_{^{162}\mathrm{Dy}}=14(1)s$) and $^{164}$Dy BEC ($\tau_{^{164}\mathrm{Dy}}=0.6(3)s$) at a magnetic field of 9.7\,G. We normalize the atom number $N$ to the atom number $N_0$ at zero hold time. For this measurement, $N_0=1.2\cdot10^4$ for $^{162}$Dy and $N_0=5\cdot10^3$ $^{164}$Dy.}
\label{fig2}
\end{figure}

The starting point of our experiment is a dual-isotope magneto-optical trap (DI-MOT) of $^{162}$Dy and $^{164}$Dy, as described in Ref. \cite{duerbeck_DIMOT_2026}. 
We create spatially overlapping MOTs of $^{162}$Dy and $^{164}$Dy, without the need of additional laser systems or optical storage times. This is achieved by using first order sidebands imprinted through phase modulation onto the 421-nm slowing light and the 626-nm trapping light. After an initial loading phase of 5\,s, the DI-MOT is compressed in 500\,ms by reducing the detuning and the optical power of the MOT beams, while increasing the magnetic field gradient. The combination of radiation pressure and gravity in the narrow-line DI-MOT leads to a spontaneous polarization of the trapped Dy isotopes \cite{Dreon_optical_cooling_2017}. As a result, both isotopes are prepared in their lowest absolute Zeeman sublevel $m_{J}=-8$ \cite{Maier_narrowline_MOT_Dy_2014, Dreon_optical_cooling_2017}, with very similar phase-space density (PSD) conditions. At the end of the compression phase we typically trap around $3\cdot10^{7}$ atoms at 10\,$\mu$K for each isotope. \newline
Atoms are then transferred into a crossed-beam optical dipole trap (ODT) for forced evaporative cooling. The trap is formed by two beams derived from a 50-W fiber laser at 1064\,nm, crossing at an angle of 45° at the DI-MOT position, each with a circular waist of 22\,$\mu$m. To increase the trapping volume, each ODT beam is spatially modulated to widen its effective horizontal waist to approximately 55\,$\mu $m. With these settings, we typically transfer around $10\%$ of the atoms from the DI-MOT into the crossed ODTs. After the loading phase, the mixture is evaporatively cooled by lowering the optical trap power through a series of exponential ramps. A key advantage of the Dy-isotope mixture is the almost identical electronic polarizability of the two isotopes at the trapping wavelength, which ensures almost equal trapping potential throughout the evaporation sequence. We find that the mixture is stable at a magnetic field around 9.7\,G. In this region, we optimize the evaporation sequence by imaging each isotope independently via resonant absorption imaging. We find that evaporation ramps of similar duration to the single-isotope case are sufficient to reach quantum degeneracy. Below the critical temperature for Bose-Einstein condensation, both isotopes develop the characteristic bimodal momentum distribution. Time-of-flight images of the two Bose-Einstein condensates (BECs) expanding from the same crossed-beam ODT are shown in Fig. \ref{fig1}c. We can tune the population imbalance between the two isotopes via the radio-frequency drive of the electro-optic modulator (EOM) that generates the 626-nm DI-MOT trapping light. As detailed in Ref. \cite{duerbeck_DIMOT_2026}, detuning this drive by $\delta_\mathrm{EOM}$ from half the isotope shift of the 626-nm transition makes the two isotopes experience different radiation pressure. This difference shifts the equilibrium position of each laser-cooled cloud (Fig. \ref{fig1}a) and makes the transfer efficiency into the ODT isotope-dependent. We exploit this effect to control the population imbalance in the mixture. We can arbitrarily prefer one isotope over the other, however, it seems easier to realize larger clouds of $^{162}$Dy than $^{164}$Dy. Our analysis suggests that $^{162}$Dy displays lower three-body losses in the magnetic field region around 9.7\,G, which results in larger degenerate $^{162}$Dy gases at the end of the sequence. For simplicity, we focus for the remainder of the paper on the condition $\delta_\mathrm{EOM}=0$, which corresponds to a similar initial loading into the ODT. Nevertheless, this tool may be useful in the future to investigate impurity physics in dipolar mixtures \cite{baroni_quantum_2024,Volosniev2023}. At $\delta_\mathrm{EOM}=0$, we typically realize a Bose-Bose mixture consisting of $N=2.1\cdot10^4$ $^{162}$Dy atoms and $N=1.3\cdot10^4$ $^{164}$Dy atoms, with a condensed fraction around 50$\%$ and 20$\%$, respectively, at an equilibrium temperature around 20\,nK. Here, we measure trap frequencies of 168(2)\,Hz, 68(2)\,Hz, and 30(1)\,Hz along the three trap axes. The tightest confinement is along the vertical axis, parallel to both gravity and the bias magnetic field (see Fig. \ref{fig1}b). Cooling even further, we can realize condensates with no measurable thermal fraction consisting of around $1.2\cdot10^4$ $^{162}$Dy atoms and $5\cdot10^3$ $^{164}$Dy atoms. \newline
To better characterize the evaporation efficiency of our dipolar mixture, we track the peak PSD $\rho_0$ as a function of the atom number $N$ throughout the evaporation sequence for both the single-isotope case and the mixture. The peak PSD $\rho_0$ is defined as $\rho_0= n\lambda_{T}^3$ with $\lambda_T$ the thermal de Broglie wavelength for an atom of mass $m$ at temperature $T$ \cite{Krstaji_Characterization_three-body_2023}. From the data shown in Fig.\ref{fig2}, we extract the evaporation efficiency $\gamma=-\frac{\partial \ln \rho_0}{\partial \ln N}$. We find almost identical values of $\gamma$ in all cases, with 2.6(5) for $^{162}$Dy and 2.7(8) for $^{164}$Dy. Importantly, in the mixture both isotopes show a comparable evaporation efficiency, $\gamma=2.4$(1.0), indicating efficient thermalization between the two species with neither acting preferentially as a coolant for the other \cite{Modugno2001, Demarco_2019}.\newline
While the mixture reaches the onset of condensation with nearly equal atom numbers in the two isotopes, $^{164}$Dy seems to suffer from stronger inelastic losses once the condensate has formed. At the end of evaporation, the $^{164}$Dy BEC shows a significantly lower lifetime with respect to a $^{162}$Dy BEC. The lifetime of both isotopes at the magnetic field $B_\mathrm{BEC}=9.7$\,G is shown in Fig. \ref{fig2}b. The rapid loss of $^{164}$Dy at short times points towards some isotope-specific three-body-loss mechanism. We point out that this mechanism may also affect the evaporation efficiency in the single-isotope case. Evaporating at $B_\mathrm{BEC}$, we can produce pure condensates of more than $1.3\cdot 10^5$ $^{162}$Dy atoms, while BECs of $^{164}$Dy are limited to $8.0\cdot 10^4$ atoms.\newline
The lower atom number in the degenerate mixture compared to the single-isotope cases originates from the smaller initial loading of the DI-MOT, which is limited by the available laser-cooling power \cite{duerbeck_DIMOT_2026}. When the population is balanced, the isotopes reach quantum degeneracy at the same point of the evaporation, confirming that the optical potentials experienced by $^{162}$Dy and $^{164}$Dy at 1064\,nm are nearly identical. At the end of evaporation, we estimate a rather small difference in gravitational sag on the order of 100\,nm.

\section{Interspecies Feshbach resonances}

\begin{figure}[b!]
 \centering
        \includegraphics[width=1\textwidth]{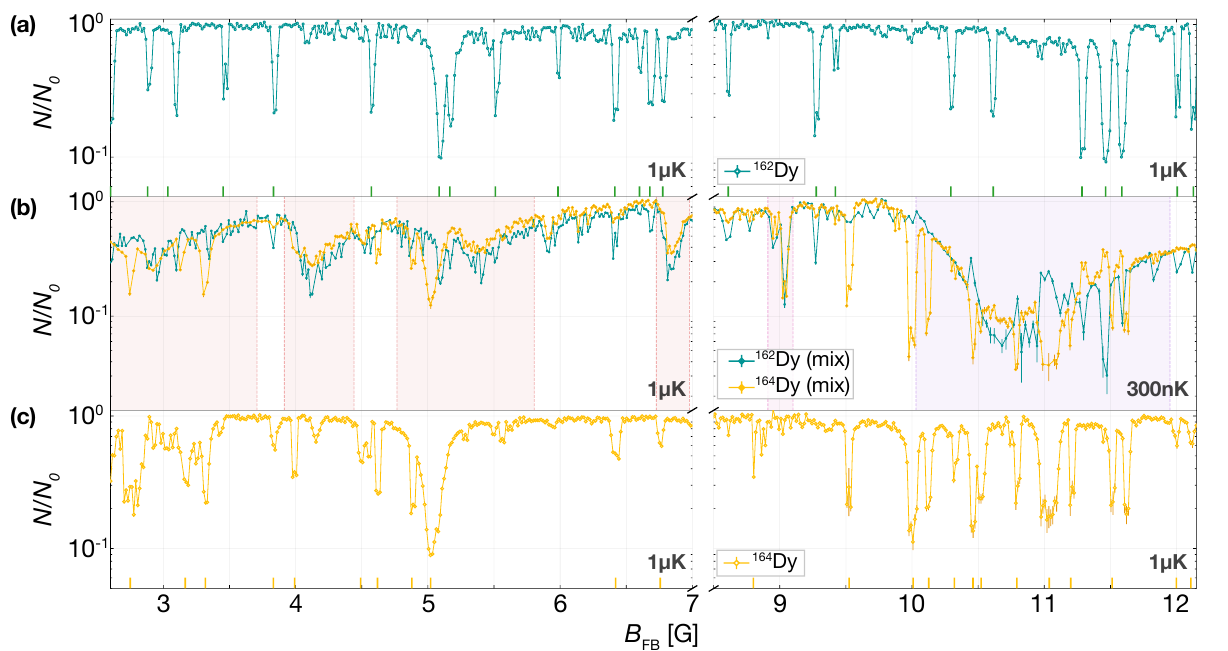}
 \caption{Interspecies Feshbach resonances between $^{162}$Dy and $^{164}$Dy. (a) \& (c) show the single-isotope atom-loss spectra over 2.6-7\,G and 8.5-12.1\,G for $^{162}$Dy (green) and $^{164}$Dy (orange), recorded at 1\,$\mu$K with a 1\,s hold time. (b) shows the atom-loss spectra of the $^{162}$Dy-$^{164}$Dy mixture over the same field range, recorded at 300\,nK with a 500\,ms hold time. Shaded regions highlight interspecies resonances. A broad resonance, extending from approximately 10.1 to 11.95\,G, is centered at $B_0 = 10.8(3)$\,G, offering the opportunity to easily control the interspecies scattering length $a_{12}$ in the Bose-Bose mixture.}
\label{fig3}
\end{figure}

Lanthanide quantum gases are well known to show a dense spectrum of Feshbach resonances \cite{chomaz_dipolar_2022, Maier_Emergence_of_Chaotic_2015, durastante2020feshbach,Frisch2014}. To asses the presence of interspecies Feshbach resonances in our isotopic mixture, we perform atom-loss spectroscopy as a function of magnetic field. We prepare mixtures at temperatures ranging from 300\,nK to 1\,$\mu$K, with atom numbers ranging from $10^5$ to $3\cdot10^5$ for each isotope. To enhance the interspecies Feshbach signal, we balance the isotope populations during the scan. The atom-number difference is kept below 10\% of the more abundant species, typically $^{162}$Dy. We then linearly ramp the magnetic field from the evaporation field $B_\mathrm{BEC}=9.7$\,G to a target field $B_{\mathrm{FB}}$ in 20\,ms, hold for 500\,ms, and measure the remaining trapped atoms $N$ via absorption imaging. For each target field $B_{\mathrm{FB}}$, we normalize $N$ to a reference value $N_0$ obtained by holding at $B_\mathrm{BEC}$ for the same duration. We calibrate the magnetic field via radio-frequency spectroscopy using an antenna inspired by the design of Ref. \cite{scazza_low-impedance_2025}. The magnetic field shows a shot-to-shot stability of around 10\,mG.\newline
To identify an interspecies Feshbach resonance, we compare the atom-loss spectra of the mixture (Fig. \ref{fig3}b) to similar atom-loss spectra of each individual isotope (Fig. \ref{fig3}a-\ref{fig3}c). Features appearing in the mixture spectrum but absent from both single-species spectra are attributed to the presence of an interspecies Feshbach resonance. In the low-field region ($B_\mathrm{FB}<7$\,G), where the individual isotopes have convenient Feshbach resonances that allow the realization of BECs and of dipolar supersolids \cite{chomaz_dipolar_2022}, we detect different interspecies Feshbach resonances with a width around 100\,mG, as shown in Fig. \ref{fig3}. The realization of quantum degenerate Bose-Bose mixture in this magnetic field region will be explored in future investigations. At magnetic fields larger than the $B_\mathrm{BEC}=9.7$\,G, we can identify a broad interspecies resonance (see Fig. \ref{fig3}). This resonance is centered around $B_{0}=10.8(3)$\,G, extending from approximately 10.1\,G to 11.95\,G. This is a relatively broad Feshbach resonance by the standards of lanthanide systems and, in combination with the possibility to create a degenerate mixture in its surroundings, it may turn into a convenient resource for easily controlling the interspecies scattering length $a_{12}$ and explore new regimes of dipolar superfluidity. We point out that in Ref. \cite{durastante2020feshbach}, a rather similar broad Feshbach resonance was found in a Dy-Er mixture. Going back to our isotopic mixture, an additional, narrower interspecies resonance is identified at $B_0 = 9.05(5)$\,G, just below $B_\mathrm{BEC}$.\newline
The BEC production field $B_\mathrm{BEC}=9.7$\,G lies in the center of a 0.3\,G plateau where the individual isotopes do not show any single-isotope Feshbach resonance. This condition might restrict the exploration of the mixture phase-diagram, since only the interspecies scattering length can be widely controlled.


\section{Controlling interspecies interactions and the miscible-immiscible transition}
\begin{figure}[b!]
 \centering
        \includegraphics[width=1\textwidth]{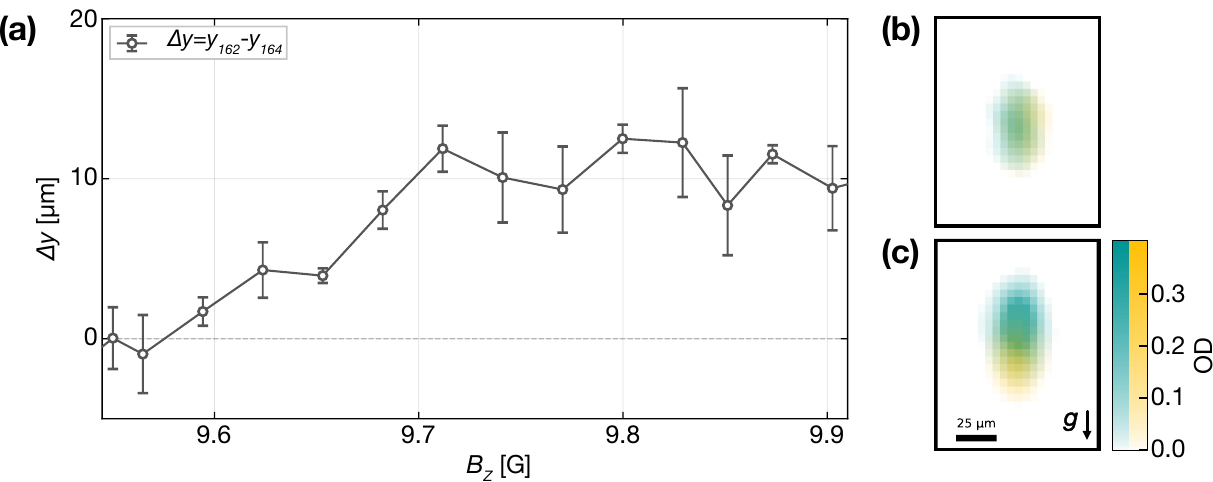}
 \caption{Evidence of miscible-immiscible transition in a Dy miture. (a) Relative vertical displacement $\Delta y$ defined as $\Delta y= y_{162}-y_{164}$, where $y_{162}$ and $y_{164}$ are the centers of mass of the $^{162}Dy$ and $^{164}Dy$ BECs, respectively, after 20\,ms time-of-flight expansion for different magnetic fields regions. (b) At a magnetic field of 9.58\,G, the two clouds are overlapped, consistent with the miscible phase. (c)At a magnetic field of 9.675\,G we move close to the resonance at 10.8(3)\,G, $a_{12}$ increases and the two clouds repel each other, consistent with an immiscible phase, where $^{162}Dy$ (green) sits on top of $^{164}Dy$ (orange). Our analysis suggests that at $B_\mathrm{BEC}$ the mixture is produced in the immiscible regime.}
\label{fig4}
\end{figure}
In this Section, we exploit the broad Feshbach resonance at $B_0 = 10.8(3)$\,G to tune the interspecies scattering length $a_{12}$ and provide first evidence of a miscible-immiscible transition in our dipolar Bose-Bose mixture. As described above, we produce a mixture of nearly pure BECs containing $1.2 \times 10^{4}$ $^{162}$Dy atoms and $5 \times 10^{3}$ $^{164}$Dy atoms. After evaporating at $B_\mathrm{BEC}$, we let the mixture equilibrate for 30\,ms, ramp the magnetic field to a target value $B_z$ between 9.55\,G and 9.9\,G in 5\,ms, and hold for a further 10\,ms. We then release the mixture and measure the center-of-mass separation of the two clouds after 20\,ms of time-of-flight at the target field. As shown in Fig.~\ref{fig4}, the two centers of mass overlap only below $B_\mathrm{BEC}$ and separate increasingly above it, with the heavier species $^{164}$Dy always sitting lower.\newline
The field-dependent center-of-mass separation provides a direct probe of the interplay between the inter- and intraspecies scattering lengths. Owing to the broad interspecies resonance, we assume that the change in magnetic field translates predominantly into a change in the interspecies scattering length $a_{12}$, though a more detailed investigation remains for future work, including for instance the effects of the narrow $^{164}$Dy resonance at 9.52 G.\newline \newline
As $a_{12}$ is tuned across the broad Feshbach resonance, the balance between inter- and intraspecies contact interactions shifts, and the resulting center-of-mass displacement tracks the approach toward the miscibility threshold. A similar qualitatively field-dependent center-of-mass separation has been used in non-dipolar dual-species condensates as a continuous probe of the miscibility parameter~\cite{Hall1998,Papp2008, McCarron2011, Wacker2015, Burchianti2018, Politi2022}, where full phase separation manifests as a strong deviation of the density distribution from the symmetry of the trapping potential~\cite{Papp2008, Wacker2015}. In our experiment, the smooth and monotonic variation of the center-of-mass separation with magnetic field is consistent with tuning the mixture across or near the miscibility crossover. This constitutes the first evidence of interaction-driven spatial reorganization in a mixture of Dy BECs. A definitive demonstration of full phase separation would require in-trap density profiles resolving the spatial segregation of the two components, which we leave to future work. In the immiscible regime, we do not observe the sharp cloud edges seen in alkali mixtures~\cite{Wacker2015,Burchianti2018}. This is likely limited by the resolution of our current imaging system. Additionally, the strong dipole-dipole interaction may further soften the edges through magnetostriction~\cite{chomaz_dipolar_2022,trautmann_dipolar_2018}. Nevertheless, the broad Feshbach resonance at 10.8\,G gives direct control of $a_{12}$ over a wide range. This makes the $^{162}$Dy-$^{164}$Dy mixture a promising platform for studying density-modulated transitions in multi-component dipolar gases. For such mixtures, the anisotropic dipole-dipole interaction is expected to reshape both the phase boundary and the structure of the phase-separated state~\cite{kumar_miscibility_2017}.

\section{Conclusions}
In summary, we have reported the first realization of a quantum degenerate Bose-Bose mixture of $^{162}$Dy and $^{164}$Dy, and started to characterize its collisional properties. We have located many interspecies Feshbach resonances in this system, most notably a broad resonance at $B_0 = 10.8(3)$\,G that provides a wide and experimentally convenient tuning range for $a_{12}$. Exploiting this resonance, we have demonstrated field-dependent control of the interspecies interactions and presented first evidence of a miscible-immiscible transition between two Dy BECs. The $^{162}$Dy-$^{164}$Dy mixture combines the large magnetic dipole moment of Dy with the tunability of a dual-species system. The tools demonstrated here lay the groundwork for future studies of dipolar supersolidity, droplet physics, and exotic density-modulated phases in multi-component quantum gases~\cite{Bisset_Quantum_droplets_2021, bland_alternating-domain_2022,AuChen2022,AuChen2021, Kirkby2024,masalaeva2026quantumvortexchannelsjosephson}.

%
%







\begin{thebibliography}{10}

\bibitem{chomaz_dipolar_2022}
L.~Chomaz, I.~Ferrier-Barbut, F.~Ferlaino, B.~Laburthe-Tolra, B.~L. Lev, and
  T.~Pfau.
\newblock Dipolar physics: A review of experiments with magnetic quantum gases.
\newblock {\em Reports on Progress in Physics}, 86(2):026401, dec 2022.

\bibitem{tanzi_observation_2019}
L.~Tanzi, E.~Lucioni, F.~Fam\`{a}, J.~Catani, A.~Fioretti, C.~Gabbanini,
  R.~N. Bisset, L.~Santos, and G.~Modugno.
\newblock Observation of a dipolar quantum gas with metastable supersolid
  properties.
\newblock {\em Phys. Rev. Lett.}, 122:130405, apr 2019.

\bibitem{bottcher_transient_2019}
F.~B\"{o}ttcher, J.-N. Schmidt, M.~Wenzel, J.~Hertkorn, M.~Guo, T.~Langen,
  and T.~Pfau.
\newblock Transient supersolid properties in an array of dipolar quantum
  droplets.
\newblock {\em Phys. Rev. X}, 9:011051, mar 2019.

\bibitem{chomaz_long-lived_2019}
L.~Chomaz, D.~Petter, P.~Ilzh\"{o}fer, G.~Natale, A.~Trautmann, C.~Politi,
  G.~Durastante, R.~M.~W. van Bijnen, A.~Patscheider, M.~Sohmen, M.~J. Mark,
  and F.~Ferlaino.
\newblock Long-lived and transient supersolid behaviors in dipolar quantum
  gases.
\newblock {\em Phys. Rev. X}, 9:021012, apr 2019.

\bibitem{trautmann_dipolar_2018}
A.~Trautmann, P.~Ilzh\"{o}fer, G.~Durastante, C.~Politi, M.~Sohmen, M.~J.
  Mark, and F.~Ferlaino.
\newblock Dipolar quantum mixtures of erbium and dysprosium atoms.
\newblock {\em Phys. Rev. Lett.}, 121:213601, nov 2018.

\bibitem{bland_alternating-domain_2022}
T.~Bland, E.~Poli, L.~A. Pe\~{n}a Ardila, L.~Santos, F.~Ferlaino, and R.~N.
  Bisset.
\newblock Alternating-domain supersolids in binary dipolar condensates.
\newblock {\em Physical Review A}, 106(5):053322, nov 2022.

\bibitem{Bisset_Quantum_droplets_2021}
R.~N. Bisset, L.~A. Pe\~{n}a Ardila, and L.~Santos.
\newblock Quantum droplets of dipolar mixtures.
\newblock {\em Phys. Rev. Lett.}, 126:025301, jan 2021.

\bibitem{scheiermann2023}
D.~Scheiermann, L.~A. Pe\~{n}a Ardila, T.~Bland, R.~N. Bisset, and
  L.~Santos.
\newblock Catalyzation of supersolidity in binary dipolar condensates.
\newblock {\em Phys. Rev. A}, 107:L021302, feb 2023.

\bibitem{Arazo2023}
M.~Arazo, A.~Gallem\'{\i}, M.~Guilleumas, R.~Mayol, and L.~Santos.
\newblock Self-bound crystals of antiparallel dipolar mixtures.
\newblock {\em Phys. Rev. Res.}, 5:043038, oct 2023.

\bibitem{AuChen2022}
A.-C. Lee, D.~Baillie, and P.~B. Blakie.
\newblock Stability of a flattened dipolar binary condensate: Emergence of the
  spin roton.
\newblock {\em Phys. Rev. Res.}, 4:033153, aug 2022.

\bibitem{Burdick_Fermionic_suppression_2015}
N.~Q. Burdick, K.~Baumann, Y.~Tang, M.~Lu, and B.~L. Lev.
\newblock Fermionic suppression of dipolar relaxation.
\newblock {\em Phys. Rev. Lett.}, 114:023201, jan 2015.

\bibitem{claudeopticalmanipulationspinstates2024}
F.~Claude, L.~Lafforgue, J.~J.~A. Houwman, M.~J. Mark, and F.~Ferlaino.
\newblock Optical manipulation of spin states in ultracold magnetic atoms via
  an inner-shell hz transition.
\newblock arXiv:2405.01499, 2024.

\bibitem{Lecomte_production_2025}
M.~Lecomte, A.~Journeaux, J.~Veschambre, J.~Dalibard, and R.~Lopes.
\newblock Production and stabilization of a spin mixture of ultracold dipolar
  bose gases.
\newblock {\em Phys. Rev. Lett.}, 134:013402, jan 2025.

\bibitem{duerbeck_DIMOT_2026}
M.~Duerbeck, L.~Reihs, J.~Seifert, B.~Choudhari, J.~P. Marulanda-Serna,
  N.~Werum, M.~De~Pas, G.~Meijer, and G.~Valtolina.
\newblock Dual-isotope narrow-line {MOT} of dysprosium by phase modulation.
\newblock {\em Submitted}, 2026.

\bibitem{Dreon_optical_cooling_2017}
D.~Dreon, L.~A. Sidorenkov, C.~Bouazza, W.~Maineult, J.~Dalibard, and
  S.~Nascimbene.
\newblock Optical cooling and trapping of highly magnetic atoms: the benefits
  of a spontaneous spin polarization.
\newblock {\em Journal of Physics B: Atomic, Molecular and Optical Physics},
  50(6):065005, mar 2017.

\bibitem{Maier_narrowline_MOT_Dy_2014}
T.~Maier, H.~Kadau, M.~Schmitt, A.~Griesmaier, and T.~Pfau.
\newblock Narrow-line magneto-optical trap for dysprosium atoms.
\newblock {\em Opt. Lett.}, 39(11):3138--3141, jun 2014.

\bibitem{Krstaji_Characterization_three-body_2023}
M.~Krstaji\'{c}, P.~Juh\'{a}sz, J.~Ku\v{c}era, L.~R. Hofer, G.~Lamb,
  A.~L. Marchant, and R.~P. Smith.
\newblock Characterization of three-body loss in $^{166}$Er and optimized
  production of large bose-einstein condensates.
\newblock {\em Phys. Rev. A}, 108:063301, dec 2023.

\bibitem{Modugno2001}
G.~Modugno, G.~Ferrari, G.~Roati, R.~J. Brecha, A.~Simoni, and M.~Inguscio.
\newblock Bose-einstein condensation of potassium atoms by sympathetic cooling.
\newblock {\em Science}, 294(5545):1320--1322, 2001.

\bibitem{Demarco_2019}
L.~De~Marco, G.~Valtolina, K.~Matsuda, W.~G. Tobias, J.~P. Covey, and
  J.~Ye.
\newblock A degenerate fermi gas of polar molecules.
\newblock {\em Science}, 363(6429):853--856, 2019.

\bibitem{baroni_quantum_2024}
C.~Baroni, G.~Lamporesi, and M.~Zaccanti.
\newblock Quantum mixtures of ultracold gases of neutral atoms.
\newblock {\em Nature Reviews Physics}, 6(12):736--752, nov 2024.

\bibitem{Volosniev2023}
A.~G. Volosniev, G.~Bighin, L.~Santos, and L.~A. Pe\~{n}a Ardila.
\newblock Non-equilibrium dynamics of dipolar polarons.
\newblock {\em SciPost Phys.}, 15:232, 2023.

\bibitem{chomaz_dipolar_2022b}
L.~Chomaz et~al.
\newblock Dipolar physics: A review of experiments with magnetic quantum gases.
\newblock {\em Reports on Progress in Physics}, 86(2):026401, dec 2022.

\bibitem{Maier_Emergence_of_Chaotic_2015}
T.~Maier, H.~Kadau, M.~Schmitt, M.~Wenzel, I.~Ferrier-Barbut, T.~Pfau,
  A.~Frisch, S.~Baier, K.~Aikawa, L.~Chomaz, M.~J. Mark, F.~Ferlaino,
  C.~Makrides, E.~Tiesinga, A.~Petrov, and S.~Kotochigova.
\newblock Emergence of chaotic scattering in ultracold er and dy.
\newblock {\em Phys. Rev. X}, 5:041029, nov 2015.

\bibitem{durastante2020feshbach}
G.~Durastante, C.~Politi, M.~Sohmen, P.~Ilzh\"{o}fer, M.~J. Mark, M.~A.
  Norcia, and F.~Ferlaino.
\newblock Feshbach resonances in an erbium-dysprosium dipolar mixture.
\newblock {\em Phys. Rev. A}, 102:033330, sep 2020.

\bibitem{Frisch2014}
A.~Frisch, M.~Mark, K.~Aikawa, F.~Ferlaino, J.~L. Bohn, C.~Makrides,
  A.~Petrov, and S.~Kotochigova.
\newblock Quantum chaos in ultracold collisions of gas-phase erbium atoms.
\newblock {\em Nature}, 507(7493):475--479, 2014.

\bibitem{scazza_low-impedance_2025}
F.~Scazza, G.~Del~Pace, L.~Pieri, R.~Concas, W.~J. Kwon, and G.~Roati.
\newblock A low-impedance radio-frequency circuit for fast spin manipulations
  in cold alkali atoms.
\newblock {\em Review of Scientific Instruments}, 96(10):104713, oct 2025.

\bibitem{Hall1998}
D.~S. Hall, M.~R. Matthews, J.~R. Ensher, C.~E. Wieman, and E.~A. Cornell.
\newblock Dynamics of component separation in a binary mixture of bose-einstein
  condensates.
\newblock {\em Phys. Rev. Lett.}, 81:1539--1542, aug 1998.

\bibitem{Papp2008}
S.~B. Papp, J.~M. Pino, and C.~E. Wieman.
\newblock Tunable miscibility in a dual-species bose-einstein condensate.
\newblock {\em Phys. Rev. Lett.}, 101:040402, jul 2008.

\bibitem{McCarron2011}
D.~J. McCarron, H.~W. Cho, D.~L. Jenkin, M.~P. K\"{o}ppinger, and S.~L.
  Cornish.
\newblock Dual-species bose-einstein condensate of $^{87}$Rb and $^{133}$Cs.
\newblock {\em Phys. Rev. A}, 84:011603(R), jul 2011.

\bibitem{Wacker2015}
L.~Wacker, N.~B. J\o{}rgensen, D.~Birkmose, R.~Horchani, W.~Ertmer,
  C.~Klempt, N.~Winter, J.~Sherson, and J.~J. Arlt.
\newblock Tunable dual-species bose-einstein condensates of $^{39}$K and
  $^{87}$Rb.
\newblock {\em Phys. Rev. A}, 92:053602, nov 2015.

\bibitem{Burchianti2018}
A.~Burchianti, C.~D'Errico, S.~Rosi, A.~Simoni, M.~Modugno, C.~Fort, and
  F.~Minardi.
\newblock Dual-species bose-einstein condensate of $^{41}$K and $^{87}$Rb in a
  hybrid trap.
\newblock {\em Phys. Rev. A}, 98:063616, dec 2018.

\bibitem{Politi2022}
C.~Politi, A.~Trautmann, P.~Ilzh\"{o}fer, G.~Durastante, M.~J. Mark,
  M.~Modugno, and F.~Ferlaino.
\newblock Interspecies interactions in an ultracold dipolar mixture.
\newblock {\em Phys. Rev. A}, 105:023304, feb 2022.

\bibitem{kumar_miscibility_2017}
R.~K. Kumar, P.~Muruganandam, L.~Tomio, and A.~Gammal.
\newblock Miscibility in coupled dipolar and non-dipolar bose--einstein
  condensates.
\newblock {\em Journal of Physics Communications}, 1(3):035012, nov 2017.

\bibitem{AuChen2021}
A.-C. Lee, D.~Baillie, P.~B. Blakie, and R.~N. Bisset.
\newblock Miscibility and stability of dipolar bosonic mixtures.
\newblock {\em Phys. Rev. A}, 103:063301, jun 2021.

\bibitem{Kirkby2024}
W.~Kirkby, A.-C. Lee, D.~Baillie, T.~Bland, F.~Ferlaino, P.~B. Blakie, and
  R.~N. Bisset.
\newblock Excitations of a binary dipolar supersolid.
\newblock {\em Phys. Rev. Lett.}, 133:103401, sep 2024.

\bibitem{masalaeva2026quantumvortexchannelsjosephson}
N.~Masalaeva, W.~Kirkby, F.~Ferlaino, and R.~N. Bisset.
\newblock Quantum vortex channels as josephson junctions.
\newblock arXiv:2602.01889, 2026.

\end{thebibliography}

\begin{acknowledgments}
We thank H. Haak for the mechanical design of our apparatus, and S. Kray, R. Thomas, D. Fontoura Barroso and the members of the electronic and mechanical workshops at FHI for their valuable technical assistance. G.V. acknowledges support from the European Union (ERC, LIRICO 101115996).
\end{acknowledgments}


\end{document}